\documentclass[lettersize,journal]{IEEEtran}
\ifCLASSINFOpdf
\else
\fi
\usepackage{amsmath,amssymb,amsfonts}
\usepackage{cleveref}
\usepackage{algorithmicx}
\usepackage{graphicx}
\usepackage{textcomp}
\usepackage{xcolor}
\usepackage{times}
\usepackage{soul}
\usepackage{url}
\usepackage[utf8]{inputenc}
\usepackage{caption}
\usepackage{booktabs}
\usepackage{algorithm}
\usepackage{algpseudocode}
\usepackage{textcomp}
\usepackage{tabularx}
\usepackage{multirow}
\usepackage{subcaption, mwe}
\newcolumntype{P}[1]{>{\centering\arraybackslash}p{#1}}
\floatname{algorithm}{Algorithm}   \algrenewcommand\algorithmicforall{\textbf{for each}}

\begin{document}
%
\title{Multi-officer Routing for Patrolling High Risk Areas Jointly Learned from Check-ins, Crime and Incident Response Data}
%
%
%

\author{Shakila~Khan~Rumi,
        Kyle~K.~Qin,
        and~Flora~D.~Salim
\thanks{Shakila Khan Rumi, Kyle K. Qin and Flora D. Salim were with the School of Computing Technologies, RMIT University, Melbourne,
VIC, Australia e-mail: (flora.salim@rmit.edu.au). \textcolor{red}{\textbf{\hl{This work has been submitted to the IEEE for possible publication. Copyright may be transferred without notice, after which this version may no longer be accessible.}}}}
}
\maketitle

\begin{abstract}
A well-crafted police patrol route design is vital in providing community safety and security in the society. Previous works have largely focused on predicting crime events with historical crime data. The usage of large-scale mobility data collected from Location-Based Social Network, or check-ins, and Point of Interests (POI) data for designing an effective police patrol is largely understudied. Given that there are multiple police officers being on duty in a real-life situation, this makes the problem more complex to solve. In this paper, we formulate the dynamic crime patrol planning problem for multiple police officers using check-ins, crime, incident response data, and POI information. We propose a joint learning and non-random optimisation method for the representation of possible solutions where multiple police officers patrol the high crime risk areas simultaneously first rather than the low crime risk areas. Later,  meta-heuristic Genetic Algorithm (GA) and Cuckoo Search (CS) are implemented to find the optimal routes. The performance of the proposed solution is verified and compared with several state-of-art methods using real-world datasets.

\end{abstract}


%

\section{Introduction}\label{sec:intro}
Public safety is an important task for the social and economical development of a country across the world. Police patrolling plays a vital role to establish the power of law and ensure the public safety. Prevention of the future crime event occurrences and prompt response to the emergency events are fundamental in effective police surveillance strategy. Limited police resources compare to the operational demands makes the problem more challenging. Moreover, funding cuts to police resources could further complicate the situation~\cite{gurdian}. Therefore, strategic patrol route planning is required to help preventing future crime events and enabling instantaneous response to emergency calls with minimal number of police officers. 

Nowadays, with the society's active participation in the web and social media, the availability of user-generated social media data is widespread. The web and big data that is either generated voluntarily by users or through a system usage contains a plethora of information about human activities and surrounding environments~\cite{kitchin2014real}. Thus, exploring web data provide unprecedented opportunities to tackle various societal problems. By harnessing this large-scale web data, the planning of patrol routes of multiple officers can be done in a more strategic and effective manner.

\begin{figure}[!ht]
    \centering
    \includegraphics[width=0.47\textwidth]{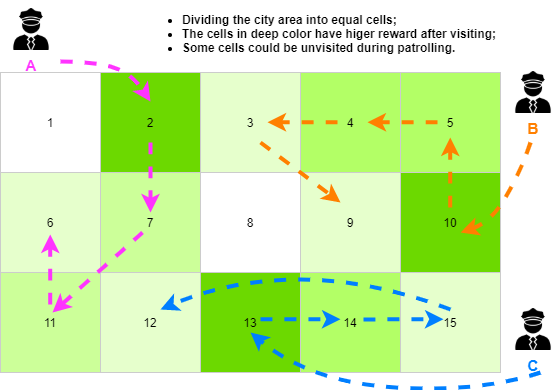}
    \caption{Demonstration of routes arrangement for three policemen in the city area that has been divided into 15 grids.}
    \label{fig:routes}
\end{figure}

Several algorithms have been developed in past to design optimal police patrol routes. Recently, the crime patrol planning problem was formulated to deal with the cooperation of a police officer for visiting different parts of city for crime event prevention and minimize the response time for emergency call based on the priority in~\cite{RumiICWSM}. However, it lacks the mutual communications between multiple police officers. For realistic solution, it is important to consider the environment, where a patrol officer can communicate with the other officers. In this paper, we extend the crime patrol planning problem for multiple police officers which aims at obtaining overall reward as much as possible. Figure~\ref{fig:routes} demonstrate the patrolling issue for three different police officers. 

This work focuses on patrol route planning problem from the outputs of dynamic crime event prediction model for short time interval. A classification model is applied to predict the crime hotspot for a time interval from heterogeneous data source. Based on the output of crime event prediction, we propose Guided Leader-based Random Keys Encoding Scheme (G-LERK) to generate the initial population of solutions which could better adapt to this special issue. One fitness function is then defined for solution optimization, which considers the prediction output, crime history and the call priority. At last, we adopt GA, CS and greedy approach with new encoding method and fitness function for optimization of multiple police patrol arrangement.

In summary, the contributions of this paper are:
\begin{enumerate}
	\item A new problem formulation for dynamic police patrolling route planning with cooperation of multiple police officers combining dynamic crime event prediction and real-time response demand.
	\item A new solution encoding method is applied to initialise population of solutions.
	\item The defined fitness function that integrates the travelling distance affect is used for acceptance criterion of global optimization.
	\item Extensive experiments using real-world data to evaluate the capability of metahuristic algorithms with different number of officers for the patrol problem.
\end{enumerate}


\section{Related Work}\label{sec:related}
Police patrol route planning include strategic solution of crime hotspot visits. Several researches have been conducted balanced and effective patrol routes design for police officers. In~\cite{chevaleyre2004theoretical}, the authors modeled the road network into a graph where the vertices were the nodes to be visited and the edges were the street segments. The weight of each edge depended on the distance between the nodes. This work aimed to reduce the idle time between two consecutive visits of a location by a police officer. This work was lack of considering the crime history in different locations. In~\cite{Chawathe2007} the weight of the edge depended on the length of the street segment and the importance of the hotspot locations. The importance of hotspot location was derived from the long-term crime event density in that location. The goal of this work was to provide optimal patrol routes which maximizes the coverage of crime hotspots in minimum distance. A cross-entropy based patrol route planning was applied in~\cite{Li2011}. The authors considered the spatial pattern of crime hotspots to suggest patrol routes and the effectiveness of collective patrol activities. In~\cite{Chen2015}, a Baysian ant colony algorithm was proposed to minimize the average idle time between two consecutive visits of crime hotspots by police officers. It used probabilistic bayesian model with ant colony algorithm which made the patrol route selections less predictable by offenders. In~\cite{xue2003decision}, the authors proposed a decision support system based on spatial choice. Here, the decision makers' preference were extracted from the incident reports.

The above described researches proposed the patrol routes based on road network and history of crime event occurrences which made the patrol environment static or change slowly with time. Dynamic environment in patrol route planning was considered in ~\cite{Chen2010} for single police officer and in ~\cite{Chen2012} for multiple police officers. The objective function in the proposed methods were reduction of the risk of potential crime rate arrival in a location in certain time interval. In~\cite{shao2018traveling}, the authors provide a dynamic probability based solution for catching cars in parking violation. Here, they applied probability based greedy and ant colony algorithm to design a route for parking officer. 

Some researchers proposed metaheuristic approaches to solve orienteering problems that are constrained by time~\cite{mei2016efficient}. But such work rarely considers predictive analytics in conjunction with optimization. Some researchers focused on path planning considering different contexts. In~\cite{rahaman2017capra}, the authors used accessibility contexts to suggest path with minimal travel time. But in this paper, the contexts are non dynamic which makes it different to our work.

None of the mentioned work optimize the patrol path generation from a proactive aspect which include prediction output in route planning. The authors considered emergency response demand and historical crime data for predicting high demand areas and used the prediction output for maximizing the demand coverage in~\cite{leigh2017predictive}. Here, the authors focused on positioning the police officers in different areas of city for demand coverage instead of route planning. 

However, uncertainties can happen in real life environment when re-planning of police patrol route will be required with existing planning. None of the mentioned works considered the fluctuation in crime event occurrences and sudden emergency call that need to be attended by a patrol officer. Our work attempts to fill the gap.
\section{Patrol Route Planning Problem}\label{sec:problem}
In this section, we discuss the patrol route planning problem in a dynamic environment. First, we introduce the patrol route planning problem for a single police officer and then we formulate the problem for multiple police officers.
\subsection{Patrol route planning for single police officer}
Here, each police officer is responsible for one police beat and moves on road network using motor vehicle with 50 km per hour average speed. The purpose of the patrolling is to prevent crime event occurrences by visiting different nodes and attend the emergency calls. G(V, E) is an indirect weighted graph where V represents the patrol nodes and E represents the edges between the nodes. The travel time between nodes represent cost, C associated with the edges between the nodes. L is a set of all nodes appeared in patrol region, $N \in V$ and emergency node, $M \in V$ which a police needs to visit. Hence, $L=N \cup M$.

Mathematically, the patrol reward for each node, $l_k$ during time interval, t  can be calculated as 
\begin{equation}\label{eq:benefit}
B(l_k(t))=exp(w_k(t))*p_k*exp(\lambda_k(t)).
\end{equation}
 Here, $w_k(t)$ and $\lambda_k(t)$ denotes the importance and crime arrival probability of node $l_k$ during time interval $t$. $w_k(t)$ depends on the density of crime event during past 3 days in node $l_k$ during time interval $t$. $\lambda_k(t)$ has been set as 0, 2, 4 for coldspot, hotspot and emergency node respectively. $p_k$ represents priority of crime. The priority of a emergency node depends on the type of response demand of the node~\footnote{http://dallascityhall.com/government/}. The priority of different nodes are noted in Table~\ref{tab:pr}. The priority of other node is set to 1. The objective is to maximize this reward for a police officer during the planning horizon,T. Mathematically, the goal is
\begin{equation}
\label{eq_o}
max \sum_{t}\sum_{l_k \in L}B(l_k(t))
\end{equation}
s.t.,\\
\begin{equation}
\label{eq_c1}
 \sum_{e_{ij}\in S, l_j \in L}C(e_{ij})+a(l_j) \leq T
\end{equation}
\begin{equation}
\label{eq_c2}
 t_a=t_c+C(e_{l,l+1})
\end{equation}
In the first constraint, Equation~\ref{eq_c1}, T represents the total duty hours of a police officer in a day. We assume that, each police officer will stay in a node for some minutes which is denoted by $a(l_j)$. According to the first constraint the total travelling time between node i to j and the staying time, in node, j can not exceed the maximum duty hours of a police officer in a day. In second constraint, Equation~\ref{eq_c2}, $t_a$ denotes the arrival time of a police officer in the destination node.
\begin{table}[h!t]
    \caption{Priority of Different node based on response type}
    \label{tab:pr}
    \centering
    \begin{tabular}{|P{0.08\textwidth}|P{0.35\textwidth}|}
    \hline
    Priority & Types of Response Call \\
    \hline
    1 & False Alarms, Nauisance Mischief, Missing Person, Missing Property, Trespass, Fraud call, Mental health, prowl \\
    \hline
    2 & {Animal complaints, Theft, Disturbances, hazards, shoplifting, property damage, suspicious circumstances} \\
    \hline
    3 & {Burglary, Liquor violations, Narcotics complaints} \\ 
    \hline
    4 & {Assaults, Sex Offender, Prostitution, Reckless burning, Robbery, Threats, Harassment} \\
    \hline
    5 & {Accident, Arrest, Homicide, Person Down/Injury, Weapons calls} \\
    \hline
    \end{tabular}
\end{table}
\subsection{Patrol route planning for multiple police officers}
In patrol route planning for multiple police officers, more than one police officer patrol a region without any area restriction. Here, the police officers need to communicate in between them for effective patrol without overlapping. An effective system is required to assign the patrol tasks to optimal police officers and to find a suitable officer to attend the emergency call on time which appears dynamically. 
Let, X is a set of police officers. Each officer $x \in X$ needs to patrol a route $r_x$ and has fixed salary $\rho$. The goal of this problem is to gain maximum patrol reward with minimum number of resource consumption. Mathematically, it can be formulated as follows:
\begin{equation}
\label{eq_3}
max \sum_{t=0}^{T}\sum_{x=1}^{|X|}\sum_{l_k \in L}B(l_k(t)) -|X|*\rho
\end{equation}
s.t.,\\
\begin{equation}
\label{eq_c4}
 \sum_{e_{ij}\in S, l_j \in L}C(e_{ij})+a(l_j) \leq T
\end{equation}
\begin{equation}
\label{eq_c5}
 \sum_{t=0}^{T}\sum_{x=1}^{|X|}\sum_{l_k \in L} x_{l_k(t)}=1
\end{equation}
Similar to the patrol route planning for single police officer, here, the first constraint, Equation~\ref{eq_c4}, represents that the total cost can not exceed total planning time, T. The second constraint, Equation~\ref{eq_c5}, denotes that a node can be assigned to only one police officer to patrol in a certain time interval of a day.

\section{DataSets}\label{sec:data}
The datasets are collected for a sector in Seattle, USA. The city is divided into total 17 different sectors. Police sector is type of spatial division which consists of more than one police beats. Each police beat is used to patrol by a police officer.
\subsubsection*{Crime Dataset}
The crime event records of Seattle, USA from ``03-2012'' -``02-2013'' are collected from from public source \footnote{https://data.seattle.gov/Public-Safety/Seattle-Police-Department-Police-Report-Incident/7ais-f98f}. The total number of crime events that happened in K police sector during this time period is 9,157.
 
 \subsubsection*{Check-in Data }
We acquire Foursquare venue data and check-in data in Seattle for the same time period as crime event records are obtained from the authors ~\cite{yang2016participatory, yang2015nationtelescope}.The venue is also known as Point Of Interest (POI). In the K police sector, the data set consists of 5,090 check-ins performed by 1,017 users a during the same time period as crime dataset.

\subsubsection*{911 Incident Response}

To simulate the emergency call to response, we use 911 Incident Response data \footnote{https://data.seattle.gov/Public-Safety/Seattle-Police-Department-911-Incident-Response/3k2p-39jp}. In K police sector, total 19,410 emergency call were required to attend by a police officer during the same time period as crime data. 

\subsubsection*{Google Maps}

The distance between two nodes are used to calculate the cost of each edge. We measure the distance between the centroid of the nodes to find the travel cost.

\newcolumntype{Y}{>{\centering\arraybackslash}X}
\section{System Approach}\label{sec:system}
\begin{figure}[!ht]
    \centering
    \includegraphics[width=0.47\textwidth]{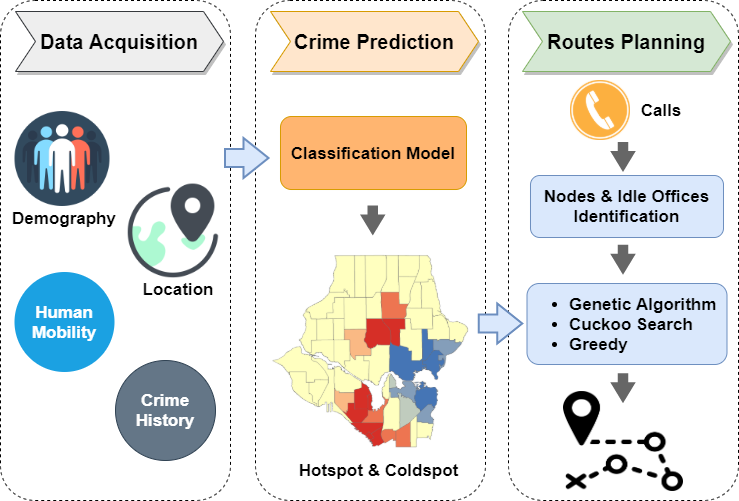}
    \caption{The framework for solving the multiple police patrol issue.}
    \label{fig:main_diagram}
\end{figure}

 The proposed system to design optimal police  patrol route consists of two parts : 1) Crime event prediction using dynamic information and 2) Patrol route generation. The framework is illustrated in Figure~\ref{fig:main_diagram}. In the following sections, we describe the crime prediction method and the prediction based dynamic route planning algorithms.
\begin{table*}[t]
	\centering
	\caption {Summary of Crime Event Prediction Features}
	\label{tbl:featurelist}
	\begin{tabularx}{\linewidth}{p{3cm}|p{4 cm}|Y} 
		\toprule
		\hline
		Feature Category & Name & Equation\\\hline
Historical Features  & $30$-days Crime Event Density & $ H_1(v_i, t)=\dfrac{\sum_{j=d-30}^{d} Cr_{j}(v_i,t)}{30}$, $Cr_j(v_i,t)$ is the number of crime events occurred at day, $j$ in node $v_i$ during time interval, $t$\\
                     \cline{2-3}
                     &  $7$-days Crime Event Density  & $ H_2(v_i, t)=\dfrac{\sum_{j=d-7}^{d} Cr_{j}(v_i,t)}{7}$  \\
                     \cline{1-3}
  & POI Distribution       &      $G_x(v_i)=\dfrac{X_x(v_i)}{X(v_i)}$     , $X_x(v_i)$ and $X(v_i)$ represents number of x-type and total venues in $v_i$ node  \\
                     \cline{2-3}
			POI based Features		 & POI Density            &  $G_2(v_i)=\frac{X(v_i)}{A(v_i)}$, Here $A(v_i)$ is the grid size                 \\
					 \cline{2-3}
					 & Location Diversity                &   $ G_3(v_i)=-\sum_{x \in X_p} (\dfrac{X_x(v_i)}{X(v_i)} \times \log \dfrac{X_x(v_i)}{X(v_i)})$, $X_p$ is total types of venues               \\
					 \cline{1-3}
 &Visitor Entropy                        & $D_1(v_i,t)=-\sum_{u \in U(v_i)}Pr_{v_i,t}(u)log_2Pr_{v_i,t}(u)$ , $U(v_i)$ denotes total social media user in $v_i$ node and $Pr_{v_i,t}$ represents the probability of giving check-in in $v_i$ during t time interval  \\
					 \cline{2-3}
                     & Visitor Homogeneity                     &     $   D_2(v_i,t)=\dfrac{\sum_{{u_1,u_2} \in U(v_i,t)}sim(u_1,u_2)_{v_i,t}}{U(v_i,t)}$, $sim(u_1, u_2)$ calculates cosine similarity between pair of users $u_1$ and $u_2$ who gave check-in in $v_i$ node during $t$ time interval       \\
                     \cline{2-3}
				Mobility based Features	 & Region Popularity                     &    $D_3(v_i,t)=\dfrac{|CH(v_i,t)|}{\sum_{v_i\in V}|CH(v_i,t)|} $, $CH(v_i,t)$ represents number of check-ins in $v_i$ node during $t$ time interval           \\
					 \cline{2-3}
					 & Visitor Ratio    &    $D_4(v_i,t)=\dfrac{|VR(v_i,t)|}{|CH(v_i,t)|}$, $VR(v_i,t)$        denotes the new users in node, $v_i$ during $t$ interval \\
					 \cline{2-3}
					 & User Count              &     Number of total users in $v_i$ node during $t$ interval, $|U(v_i,t)|$           \\
					 \cline{2-3}
					 & Observation Frequency      &   Number of total check-ins in $v_i$ node during $t$ interval,  $|CH(v_i,t)|$            \\ 
					 \hline
	\end{tabularx}
\end{table*} 

\subsection{Crime Event Prediction}
\label{sec:event_prediction}
Crime event prediction component aims to predict crime event in a short-term interval, which helps the police to design a patrol strategy in advance. With an aim to predict crime event with fine temporal granularity, we predict crime for 2 hours. Similarly, it can be predicted for other temporal granularities also. The prediction method consists of the following steps:
\subsubsection{Feature Extraction}
 Crime event prediction is a complex problem which depends upon different types of information regarding the past history of crime events, the geography of a region, and the human mobility in a region at different time intervals. We extract several features as predictors of crime event prediction model including historical features, geographic features and dynamic features.
 
Crime events are more likely to happen in the vicinity of past crime events. To retain the historical knowledge about crime event occurrence in a patrol node, we calculate features based on the crime event density. There is correlation between the type of land usages and crime event rates in a region~\cite{wo2019mixed}. The POI based features describe the geographic information of different nodes in patrol area. This type of features are extracted from foursquare POI. 

 According to the environmental criminal theories, activity location, daily life style are dominant in prompting crime event occurrences~\cite{cohen1979social, brantingham1993environment, miller2005measurement}. Human movement in a city helps us to understand the activity spaces and life style of urban people in timely manner. Now a days, booming of social media opens the door to analyze human movement in large scale. Therefore, integrating the environmental criminal theories with mobility data from social media is able to build an efficient crime event prediction model. In~\cite{rumi2018}, the authors proposed several mobility based features from  foursquare and showed the superior performance of this type of features. We apply similar dynamic features into our model.
 For node $v_i$ at time interval $t$, the training and test data includes the features which are summarized in Table \ref{tbl:featurelist}. 

 \subsubsection{Data Sampling}
Crime does not happen frequently. Therefore, many regions in many time intervals have ``no-crime'' class. With very low number of ``crime'' class, the distribution of class variables becomes asymmetrical. The classifiers with imbalanced training data cause bias towards the majority class in test result. To mitigate the problem, we balance the training set using under-sampling technique. This technique removes some classes with major distribution. In the balanced training set, the dataset contains 50\% crime data and 50\% no crime data.
\begin{figure*}[!htb]
    \centering
    \includegraphics[width=0.95\textwidth]{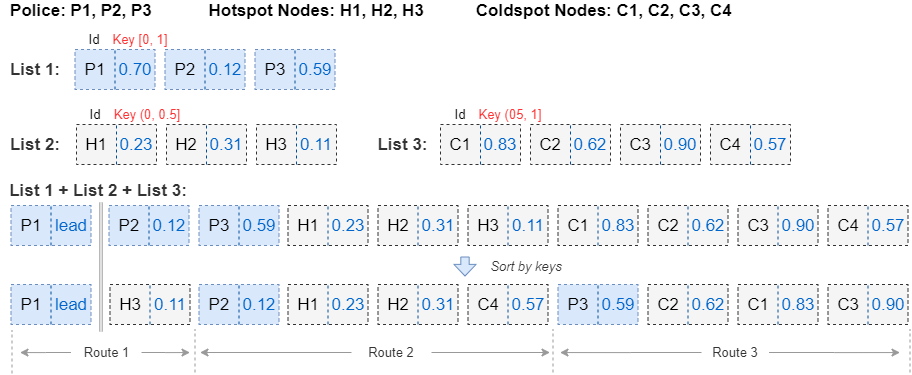}
    \caption{The illustration of new solution encoding scheme, G-LERK for multiple police patrol issue.}
    \label{fig:encoding_scheme}
\end{figure*}
\subsubsection{Classification} 
In final step, we build a classifier based on Random Forest (RF) algorithm for dynamic crime event prediction. \emph{RF} is an ensemble learning method which constructs multiple decision trees ($n_{tree}$) in a random subspace of the feature space~\cite{ho1995random}. For each subspace, the unpruned tree generates their classifications and in the final step, all the decisions generated by $n_{tree}$ are combined for final prediction~\cite{liaw2002classification}. The reason behind choosing RF as a classifier algorithm is that the convergence of the algorithm depends on only the number strong features without taking account the less important or noisy features~\cite{biau2012analysis}. The non-parametric nature of this algorithm makes it more suitable learning algorithm when the feature space is heterogeneous and multidimensional like our problem~\cite{Kadar2018}. Again, RF provided consistent and good performance in crime prediction over linear and kernel based method  in previous studies~\cite{rumi2018theft,Bogomolov2014, bogomolov2015moves}.

 \subsection{Patrol Route Planning for Multiple Police Officers}
 Here, we describe the patrol route planning framework for multiple police officers based on meta-heuristic algorithms like Genetic Algorithm (GA) and Cuckoo Search (CS). We assume that the system has real time record of emergency call, location and working status of police officers. We apply non-random, guided population generation scheme based on Leader-based Random Keys Encoding Scheme to generate the initial solution for multiple police officers. Later, we apply greedy based sorting approach to find the local optima. Finally, we apply metaheuristic algorithms, GA and CS to obtain the global optimal solution among initial population. The fitness function of the metaheuristics is defined based on the crime event density, prediction output or emergencies and priority. The fitness of a solution also depends on the arrival time of a police officer in a node.
\subsubsection{Guided Leader-based Random Keys Encoding Scheme (G-LERK)}
Our main challenge in patrol route planning for police officer is that the states of locations change with time and the police officers may prevent more crime if they visit the hotspot node first than the coldspot node. Therefore, we need to construct a temporary path for police officers which will guide them to the hotspot locations before visiting the coldspot node. We introduce a new population generation approach G-LERK based on the prediction outcome and Leader based Random Keys Encoding Scheme (LERK), which generate intial solutions for the police officers. LERK is a method which has been used to generate random ids for parking officers and parking nodes for initial solution~\cite{MTOP2019}. In LERK, the parking nodes are assigned randomly to the police officers for next task. However, in our scenario, using this encoding scheme directly in population generation might limit the chances to prevent crime. With this method, all the hotspot nodes will be assigned to first few officers or one officer. Therefore, it lowers the chances to prevent crime. To overcome this drawback, we propose G-LERK, where all idle officers start their patrol with a hotspot node instead a coldspot. Fig. \ref{fig:encoding_scheme} shows the encoding process of a solution, G-LERK for multiple police patrol issue.
In our problem, each police officer and node has unique id. The LERK combines theses ids in a link-list with a random number between 0 to 1. From, the sorted link list, the initial patrol route for each police officer is determined. In G-LERK, first, we consider two separate link lists depending on hotspot nodes and coldspot nodes. We assign different range of numbers for the hotspot nodes and colspot nodes. The first link-list combines the police officers and the hotspot nodes with a random number drawn from (0.5,1]. Later, we sort them ensuring that a police officer is in the front of the list. Similarly, we generate and sort another link list with similar police officers and coldspot nodes with a random number drawn from (0,0.5]. Finally, we combine these two lists in a single list based on the officers' id to encode a initial solution. 

\subsubsection{Local Optimization}
The local optimization helps to find the optimal solution of each sub-tour for every police officer in a new solution. We apply greedy based approach here based on the location of police officer. The node with highest benefit in a will be selected and moved to the front position of a sub-tour for the corresponding police officer. This method ensures that a police officer can have high opportunity to prevent future crime and response the demand at right time.

\subsubsection{Global Optimization}
We search for the best solution for multiple police officer patrol route problem using metaheuristic algorithms like GA, CS and greedy. For finding the optimal solution for patrol node assignment, a fitness function is defined for evaluation of the quality of solution. 
\begin{table}[h!t]
	\caption {Benefit calculation based on arrival time of police}
	\label{tab:ben}
	\centering
	\noindent
\begin{tabular}{|P{0.12\textwidth}|P{0.04\textwidth}|P{0.04\textwidth}|P{0.04\textwidth}|P{0.04\textwidth}|P{0.04\textwidth}|}
	\hline
	\multirow{2}{0.12\textwidth}{Arrival Time(min)}& \multicolumn{5}{c|}{Priority of Crime Type}\\\cline{2-6}
	 &\textbf{5}&\textbf{4}&\textbf{3}&\textbf{2}&\textbf{1}\\\hline
	 $<15$& 1&1&1&1&1\\ \hline
       [15 - 30]&0&0&0.8&0.8&0.8\\ \hline
       [31 - 60)&0&0&0&0.6&0.6\\ \hline
       $>60$ &0&0&0&0&0.5\\ \hline
	\end{tabular}
\end{table}

\textbf{Fitness Function}
In metaheuristic algorithms, in each iteration the solution is updated based on the superiority of the solution. In our experiment, a solution replaces the old one, if it achieves higher fitness value. The fitness function is defined as follows:
\begin{equation}
\label{eq:c_8}
    \sum_{r \in R} \sum_{l_k \in r} B(l_k(t))*Pr_{l_k,l_{k+1}},
\end{equation}
s.t.,\\
\begin{equation}
\label{eq:c_9}
     Pr_{l_k,l_{k+1}} \in {0,1}
\end{equation}
In Equation~\ref{eq:c_8}, R refers to the sub-routes generated for each police officer in the solution. $Pr_{l_k,l_{k+1}}$ determines the time-dependent benefit that a police officer can obtain based on the arrival time of a police officer in a node. The benefit value calculation based on the arrival time of a police officer is summarized in Table~\ref{tab:ben}. The priority value for each crime type is based on the call priority system established by Dallas police, USA~\cite{dallaspol}. 

\textbf{Genetic Algorithm (GA) for Optimization}
GA is a popular metaheuristic algorithm to identify optimal solutions for problem~\cite{whitley1994genetic}. In our approach, we use G-LERK for initial population generation for GA. Each individual in population is called chromosome. After generating the initial population, the chromosomes are ranked based on their fitness value. It separates the chromosomes into three parts including, elitist, crossover part and mutate part. Crossover and mutate part are used to generate new off-springs and elitist is kept exactly in new population. The proposed G-LERK-GA is illustrated in Algorithm~\ref{Alg:GA}.   
\begin{algorithm}  
\caption{Guided Genetic Algorithm (G-LERK-GA)}
\label{Alg:GA}
\begin{algorithmic}[1] 
\vspace{1mm}
\Require 
\Statex Set of nodes for patrol $L$; set of idle officers $P_f$; portion of population for Elitist $elitistRate$; size of population for Crossover $crossoverSize$; portion of population for Immigration $mutateRate$
\Ensure
    \Statex The optimal solution $S_{best}$
\State Initialize a population of chromosomes $S$ via the G-LERK
\While{($i <$ $maxIteration$)}
\State Rank $S$ by fitness in descending order
    \State $S_{new} \gets$ $S$
    \State Keep $elitistRate$ portion of best $S_{new}$ unchanged
    \For{($j <$ $crossoverSize$)}
     \State Select $S_{Mom}$, $S_{Dad}$ from $S$ at random
     \State Get $S_{child}$ by new crossover on $S_{Mom}$, $S_{Dad}$
     \State Apply local optimization on $S_{child}$
     \State Replace $S_i \in S_{new}$ by $S_{child}$
\EndFor
\State Immigration on a $mutateRate$ portion of worst $S_{new}$
\State $S \gets S_{new}$
\EndWhile
\State Return $S_{best}$ from $S$
\end{algorithmic} 
\end{algorithm}

\textbf{Cuckoo Search (CS) for Optimization}
CS is another popular metaheuristic algorithm, which was developed based on the parasitic behaviour of some cuckoo birds~\cite{yang2009cuckoo} and showed efficiency in several optimization problems. Similar to GA, we apply G-LERK method to  generate initial solution (population of nest), in our proposed CS algorithm. A parameter, \textit{preFly}, represents the superior candidates (cuckoos) in population. These cuckoos are allowed to fly and find high quality nests. The portion of \textit{preFly} nest, $p_c$ is determined via \textit{Levy flights}. In each iteration, the fitness value between one candidate from \textit{preFly} and another randomly selected competitor from population nest is compared for replacement. If the new solution is better than the old one, it replaces the previous one. After end of the comparison a portion of worst nest is abandoned and new nest is built. The detail of the G-LERK-CS is illustrated in Algorithm~\ref{Alg:CS}.
\begin{algorithm}  
\caption{Guided Cuckoo Search (G-LERK-CS)}
\label{Alg:CS}
\begin{algorithmic}[1] 
\vspace{1mm}
\Require 
\Statex Set of nodes for patrol $L$; set of idle officers $P_f$; number of superior candidates $preFly$; portion of population for incubation $p_c$ and drop $p_a$
\Ensure
    \Statex The optimal solution $S_{best}$
\State Initialize a population of nests $S$ via the G-LERK
\While{($i <$ $maxIteration$)}
    \For{($j <$ $preFly$)}
     \State Select a nest $S_j$ from a portion $p_c$ of top $S$
     \State Get a candidate $S_j^{'}$ based on $S_j$ via Levy flights
       \State Apply local optimization on $S_j^{'}$
     \State Evaluate $S_j^{'}$ fitness $F_j$
     \State Pick competitor $S_k \in S$ randomly with fitness $F_k$
     \If{($F_i > F_k$)}
     \State replace $S_k$ by $S_j^{'}$
     \EndIf
\EndFor
\State Abandon a portion $p_a$ of worse nests and create new ones to replace
\State Rank $S$ by fitness in descending order
\EndWhile
\State Return $S_{best}$ from $S$ 
\end{algorithmic} 
\end{algorithm}

\begin{algorithm}  
\caption{Greedy Algorithm by Importance (Imp-Greedy)}
\label{Alg:greedy}
\begin{algorithmic}[1] 
\vspace{1mm}
\Require 
\Statex Set of nodes for patrol $L$; set of idle officers $P_f$
\Ensure
    \Statex The optimal solution $S_{best}$
    \For{node $v_i \in L$}
     \State $ P_{best}$ $\gets$ null
     \State $Pr_{max}$ $\gets$ null
     \For{node $p_f \in P_f$ }
     \State $Pr$ $\gets$ Fitness($v_i$, $p_f$)
     \If($Pr > Pr_{max}$)
     \State $Pr_{max}$ $\gets$ Pr
     \State $ P_{best}$ $\gets p_f$
     \EndIf
\EndFor
\State Assign $v_i$ to $P_{best}$
\EndFor
\end{algorithmic} 
\end{algorithm}
\textbf{Greedy Algorithm for Optimization}
We also implement the greedy algorithm for arranging routes for police officers, which is simple and usually effective~\cite{akccay2007greedy,ruiz2007simple}. Two variants of greedy algorithm called Imp-Greedy and Dis-Greedy are developed in our implementation. In each arrangement of Imp-Greedy, we first calculate the essential of each node in emergency according to current position of idle officers, then a node will be allocated to the officer who has the highest essential for it. This algorithm is described in Algorithm~\ref{Alg:greedy}. And the way to calculate the essential of a node is below:
\begin{equation}
\label{eq:c_8}
    B(l_k(t))*Pr_{l_k,l_{k+1}},
\end{equation}
In contrast, instead of computing the essential for nodes, we calculate the walking or driving distance between each node and idle officers in the Dis-Greedy where a node will be assigned to the person who has the shortest distance to it.

\begin{figure*}[!ht]
    \centering
    \begin{subfigure}[b]{0.3\textwidth}
        \includegraphics[width=1\textwidth]{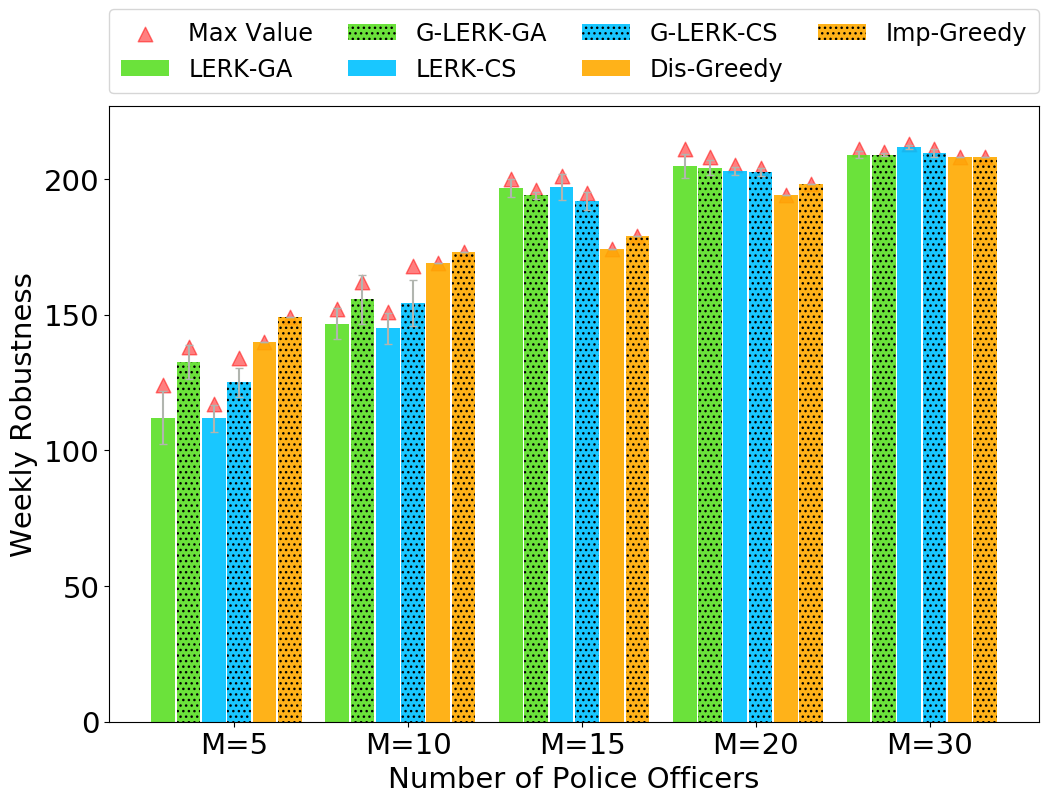}
        \caption{Week 1}
        \label{fig:week_6_13_mean_total_robustness}
    \end{subfigure}
    \quad
    \begin{subfigure}[b]{0.3\textwidth}
        \includegraphics[width=1\textwidth]{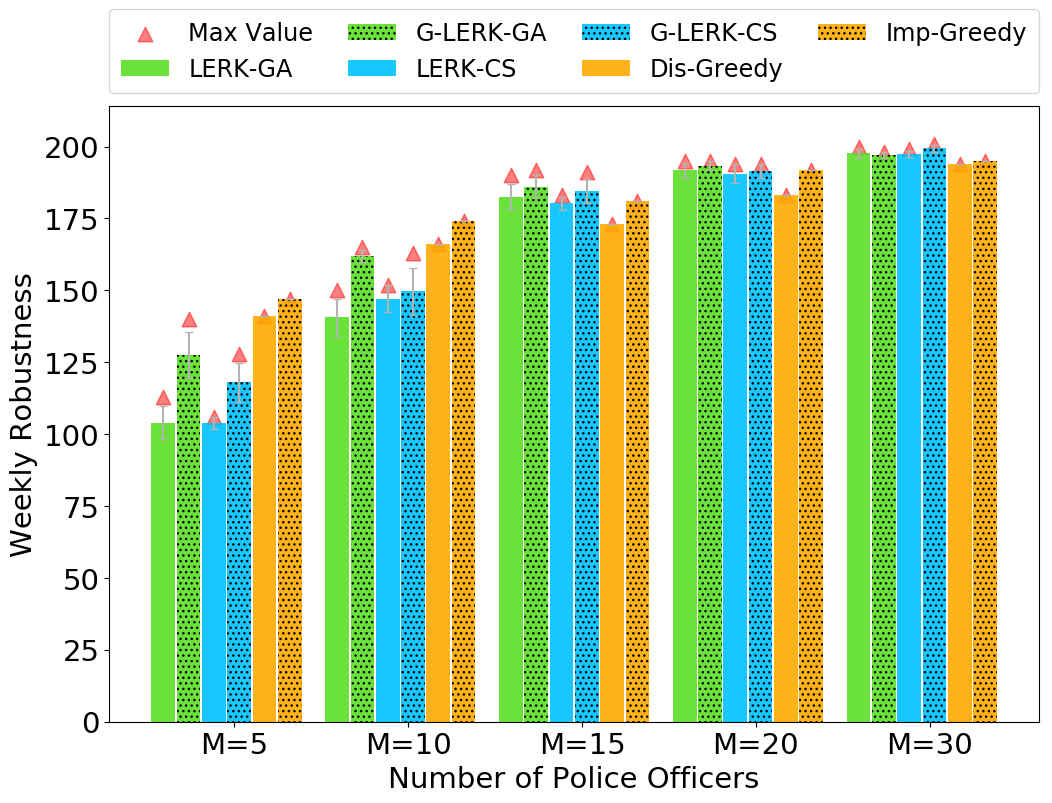}
        \caption{Week 3}
        \label{fig:week_20_27_mean_total_robustness}
    \end{subfigure}
    \quad
    \begin{subfigure}[b]{0.3\textwidth}
        \includegraphics[width=1\textwidth]{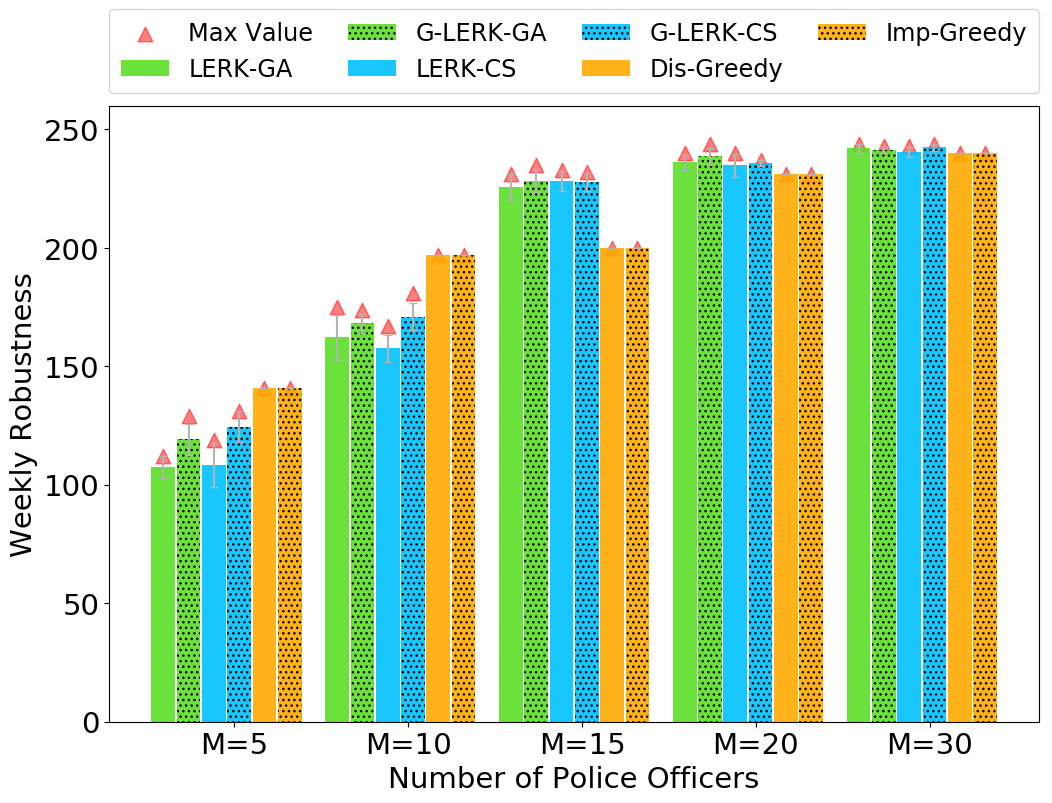}
        \caption{Week 5}
        \label{fig:week_34_41_mean_total_robustness}
    \end{subfigure}
    \caption{Weekly robustness obtained  by algorithms in five runs.}
    \label{fig:week_mean_total_robustness}
\end{figure*}

\begin{figure*}[!ht]
    \centering
    \begin{subfigure}[b]{0.3\textwidth}
        \includegraphics[width=1\textwidth]{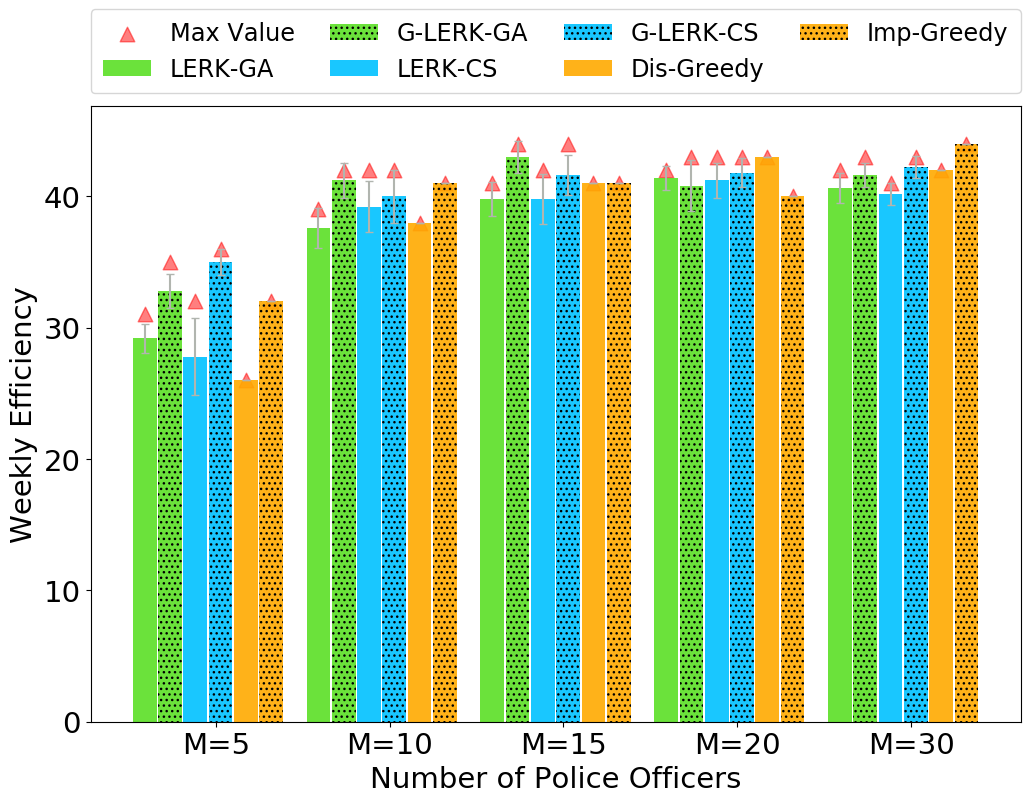}
        \caption{Week 1}
        \label{fig:week_6_13_mean_total_efficiency}
    \end{subfigure}
    \quad
    \begin{subfigure}[b]{0.3\textwidth}
        \includegraphics[width=1\textwidth]{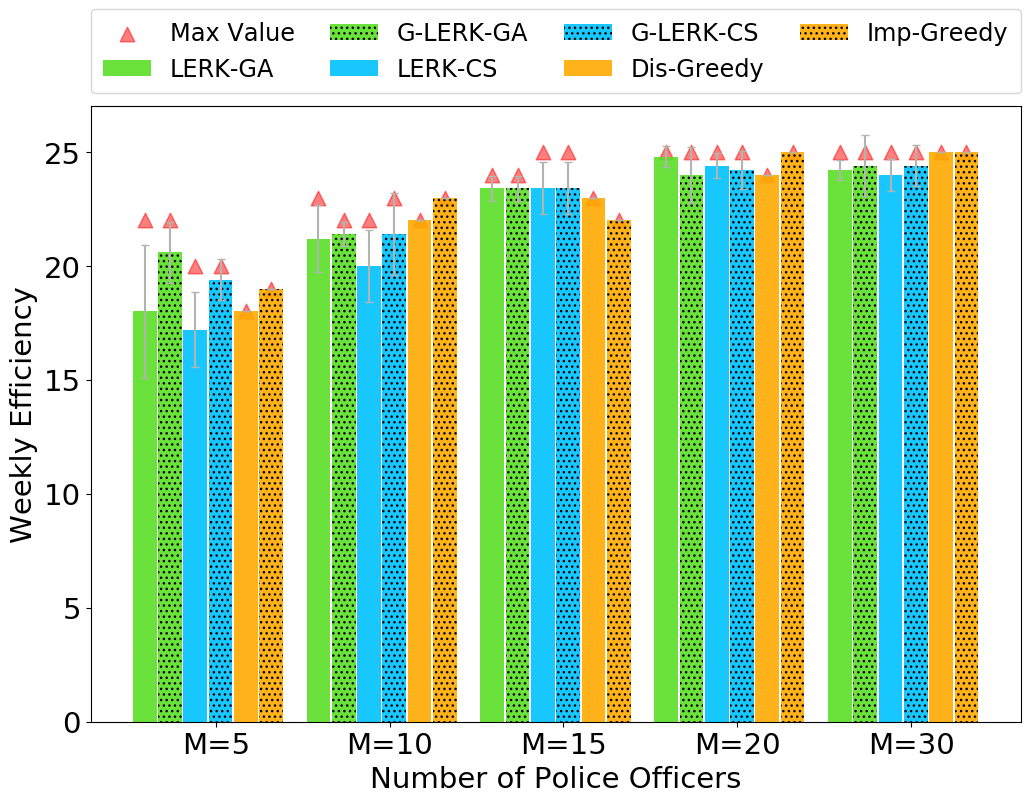}
        \caption{Week 3}
        \label{fig:week_20_27_mean_total_efficiency}
    \end{subfigure}
    \quad
    \begin{subfigure}[b]{0.3\textwidth}
        \includegraphics[width=1\textwidth]{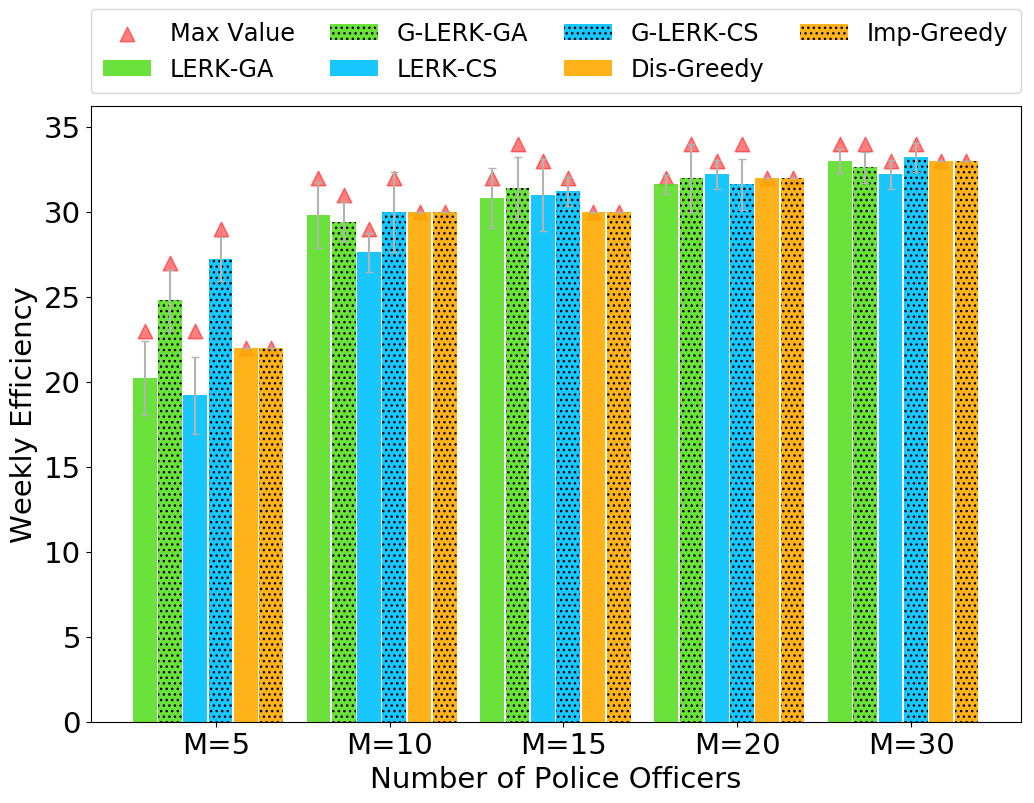}
        \caption{Week 5}
        \label{fig:week_34_41_mean_total_efficiency}
    \end{subfigure}
    \caption{Weekly efficiency obtained by algorithms in five runs.}
    \label{fig:week_mean_total_efficiency}
\end{figure*}
\section{Experiments}\label{sec:test}
The different optimisation algorithms mentioned in previous section for patrol route generation are evaluated here with real-world data described in Section~\ref{sec:data}. For the crime event prediction, a day is partitioned into intervals of 2 hours. We have used data that lies into ``04/2012 - 12/2012'' for training purpose. The data lies in ``01/2013 - 02/2013'' have been used to evaluate prediction performance and design the patrol route. The total number of emergencies and crime event occurrences of every week during patrol route planning period are given in Table \ref{tbl:total_robust_effi}. 
\begin{table}[ht]
	\centering
	\caption {Total Emergencies and Crime Occurrences of January and February in 2013.}
	\label{tbl:total_robust_effi}
	\begin{tabularx}{0.5\textwidth}{lcc} 
		\toprule
        Data Time & Total Emergencies &  Total Crime Events \\
        \midrule
        \textbf{Week 1} (07/01 - 13/01) & 215 & 48 \\
        \textbf{Week 2} (14/01 - 20/01) & 189 & 51 \\
        \textbf{Week 3} (21/01 - 27/01) & 203 & 33 \\
        \textbf{Week 4} (28/01 - 03/02) & 219 & 54 \\
        \textbf{Week 5} (04/02 - 10/02) & 251 & 45 \\
        \textbf{Week 6} (11/02 - 17/02) & 230 & 52 \\
        \bottomrule       
	\end{tabularx}
\end{table}
The experimental results for patrol route planning with system settings and evaluation methods are described in the following sections.
\subsection{System Settings}
A simulation system have been built for the experiments. Firstly, we divided the map into 94 grids equally and the initial positions of all the police officers start from the first grid. It is assumed that the polices work from 8 am to 8 pm each day to patrol around the whole map. When next patrolling duration of a police to a grid exceeds 8 pm, the system will stop the work of this police in a day. In addition, if one grid is visiting by a police, the status of both police and grid will be hided from the process of following assignments until the visit is finished. In this system, the street walking speed of officers is approximately 1.2 m/s and the frequency to scan current emergency calls, hotspot records and available polices is 1 minute. The probability to conduct the local optimization for CS and GA is 0.5,which sorts the nodes on a temporary path by their importance. 
\begin{table}[!ht]
	\centering
	\caption {Parametric configuration for LERK-CS and G-LERK-CS.}
	\label{tbl:para_cs}
	\begin{tabularx}{0.5\textwidth}{lll} 
		\toprule
		Parameter & Value & Description\\
		\midrule
          $p_a$ & 0.3 & Portion of worse nests \\
          $p_c$ & 0.6 & Portion of superior nests \\
          $\alpha$ & 0.05 & Scaling factor of step size \\
          $\lambda$ & 1 & Exponent of a power-law distribution \\
          preFly & 0.3 & Portion of superior cuckoos to seek high-quality nests \\
        \bottomrule
	\end{tabularx}
\end{table} 

\begin{table}[!ht]
	\centering
	\caption {Parametric configuration for LERK-GA and G-LERK-GA.}
	\label{tbl:para_ga}
	\begin{tabularx}{0.5\textwidth}{lll} 
		\toprule
		Parameter & Value & Description\\
		\midrule
          elitistRate & 0.2 & Proportion of population for reproduction operation \\
          crossRate & 0.3 & Proportion of population for crossover operation \\
          mutateRate & 0.2 & Proportion of population for immigration operation\\
        \bottomrule
	\end{tabularx}
\end{table}
Moreover, the population size and the number of iterations for all the implemented GA and CS are 100 and 300, respectively. The other parametric settings of these two types of algorithms are given in Tables \ref{tbl:para_cs} and \ref{tbl:para_ga}. Finally, the system was implemented in Python 3.6 and executed on a desktop with Intel i9-9900k CPU 8 Cores 3.6 GHz and Memory 64 GB DDR3. The details of experimental results are discussed in the section \ref{sec:experimental}.

\subsection{Comparison Algorithms}\label{sec:result_compa}
For comparison, we adopt random initial population generation method and different fitness function for solving the patrol route planning problem. We apply LERK-GA and LERK-CS, proposed in~\cite{MTOP2019} for comparison. Here, the authors used Random Keys Encoding Scheme, LERK to generate the initial solution. Then, we conduct experiments for the comparison between the proposed methods (G-LERK-GA and G-LERK-CS) and the original ones (LERK-GA and LERK-CS). In addition, we also implement two different versions of greedy algorithms with either prediction output or pure travelling distance consideration. They are mentioned as Dis-Greedy and Imp-Greedy.

\subsection{Evaluation Methodology}
We evaluate the performance of routes arrangement with two different metrics: \textit{efficiency} and \textit{robustness}.

\textbf{Efficiency}
 The efficiency measures how many real crime events are prevented in duty hour of a police officer. The efficiency of a patrol route is measured as follows:
\begin{equation}\label{eq:ev1}
Ef(T)=\frac{\sum_{t=t_c - 60,n_i\in N}^{t_c+60} visit (n_i)}{|N|}
\end{equation}
Here, $t_c$ represents the occurrence time of crime event; $N$ represents the nodes where the crime event occurred. $visit (n_i)$ returns 1 if  the crime node is visited by a police officer in between before or after 1-hour of occurrence.

\textbf{Robustness}
 The robustness determines how many emergencies are attended by a police officer. It also depends on the arrival time of a police officer and crime priority. If a police officer arrives to a crime spot in between 15 minutes of occurrence time the robustness has been set to 1 for all type of calls. When arrival time is between 15 - 30 minutes, it is set to 1 when the priority of call is below 4. When the arrival time increases to 31 - 60 minutes, only the calls with 1 and 2 priority returns value. If the priority of call is 1, visit of a police officer to a crime event occurrence spot after 1 hour is valuable. The robustness value is determined based on the arrival duration after making the call and the priority of the crime type. 

 \begin{figure*}[!ht]
    \centering
    \begin{subfigure}[b]{0.45\textwidth}
        \includegraphics[width=1\textwidth]{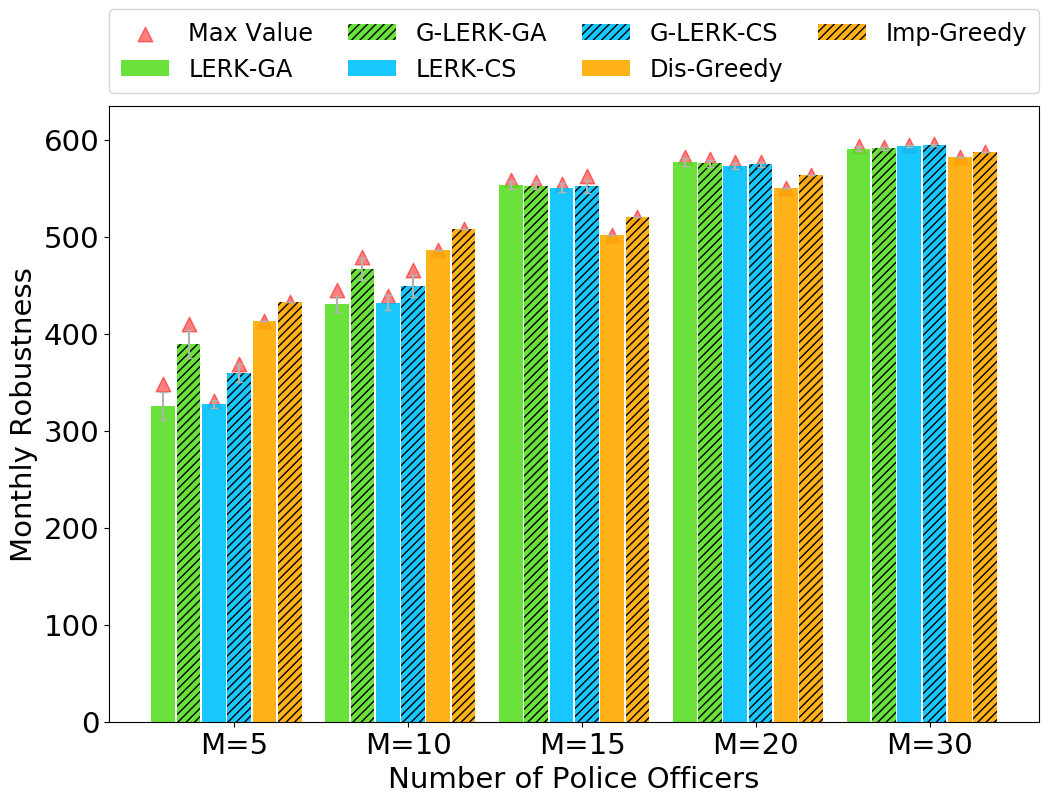}
        \caption{January}
        \label{fig:month_6_27_mean_total_robustness}
    \end{subfigure}
    \quad
    \begin{subfigure}[b]{0.45\textwidth}
        \includegraphics[width=1\textwidth]{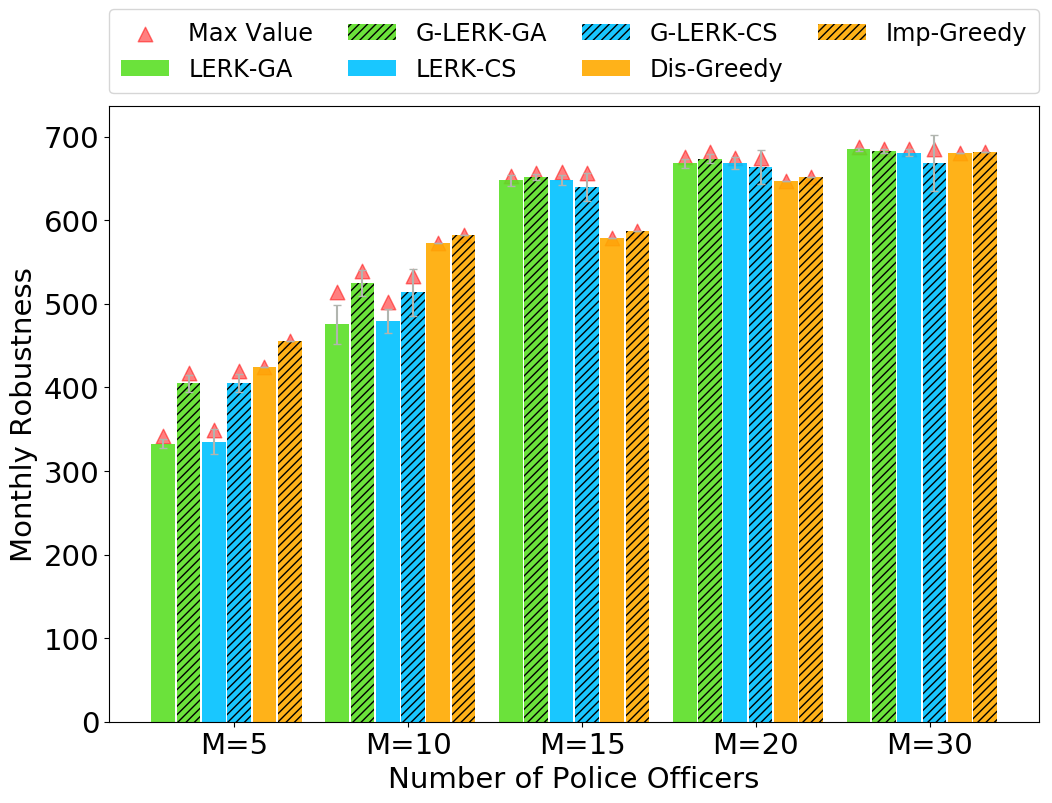}
        \caption{February}
        \label{fig:month_27_48_mean_total_robustness}
    \end{subfigure}
    \caption{Monthly robustness obtained by algorithms in five runs.}
    \label{fig:month_mean_total_robustness}
\end{figure*}

\begin{figure*}[!ht]
    \centering
    \begin{subfigure}[b]{0.45\textwidth}
        \includegraphics[width=1\textwidth]{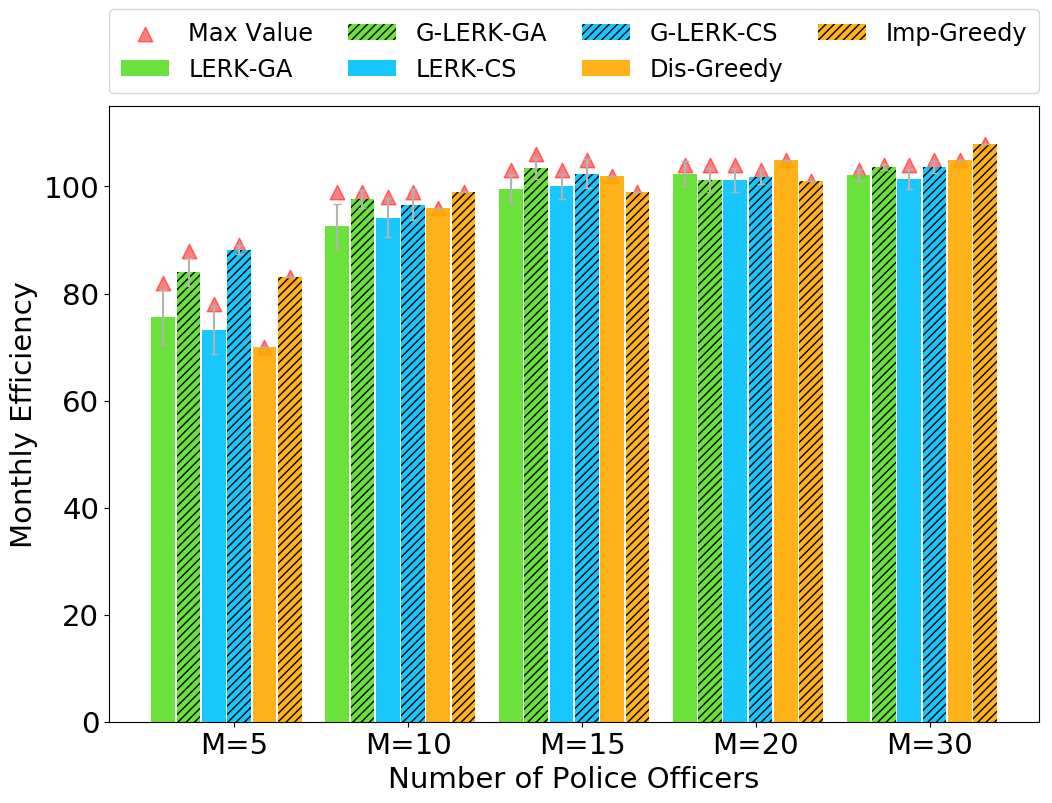}
        \caption{January}
        \label{fig:month_6_27_mean_total_efficiency}
    \end{subfigure}
    \quad
    \begin{subfigure}[b]{0.45\textwidth}
        \includegraphics[width=1\textwidth]{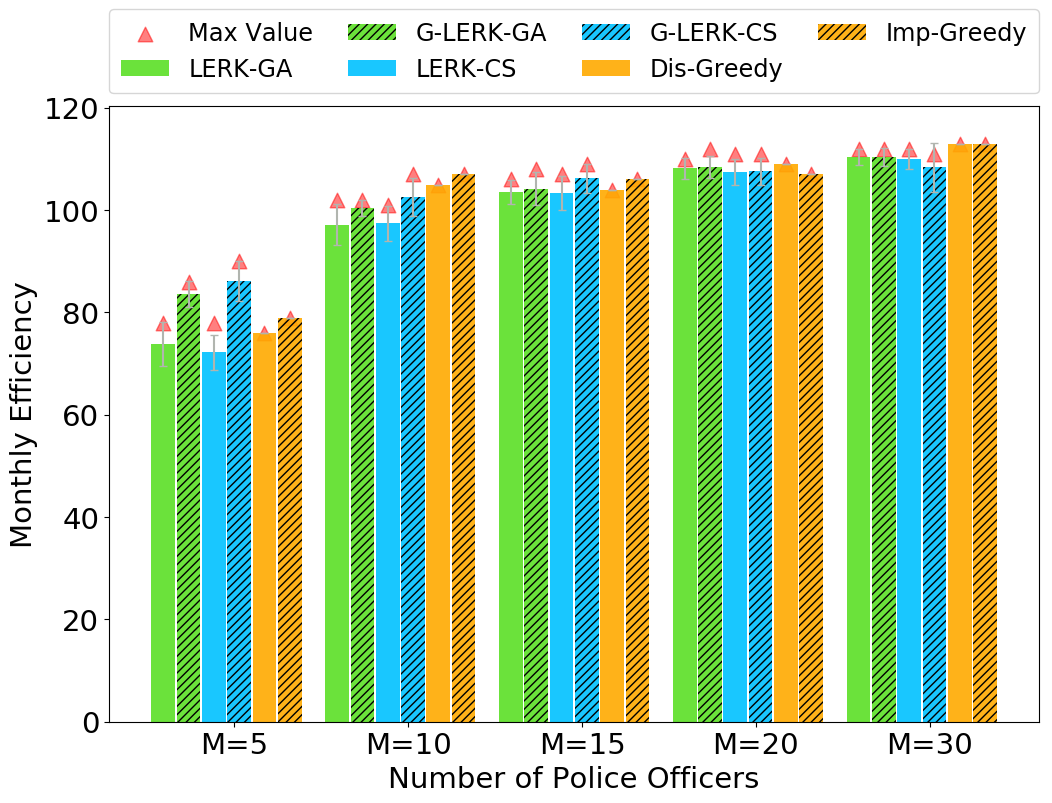}
        \caption{February}
        \label{fig:month_27_48_mean_total_efficiency}
    \end{subfigure}
    \caption{Monthly efficiency obtained by algorithms in five runs.}
    \label{fig:month_mean_total_efficiency}
\end{figure*}
\subsection{Experimental Results}
\label{sec:experimental}
In this section, we first conducted the prediction of criminal information based on different features that mentioned in Section \ref{sec:event_prediction}. Along with the emergencies and criminal events of three weeks between January and February 2013, we then evaluate the \textit{efficiency} and \textit{robustness} for six different algorithms. The detail of the weekly events is showed in Table \ref{tbl:total_robust_effi}. In these algorithms, G-LERK-CS and G-LERK-GA leverage the fitness evaluation of the destinations along with the prediction data for arranging routes of polices on map, but LERK-CS and LERK-GA only apply the fitness evaluation without considering hotspot or coldspot prediction. Greedy algorithms allocate tasks based on either the maximum essential or nearest travelling distance between offices and destinations.

\subsubsection{The Total Robustness Obtained}
In this experiment, we use 5-30 police officers to patrol the whole area and count the total \textit{robustness} that received in weekly and monthly respectively. The mean and standard deviation of the total \textit{robustness} are calculated based on five trials for each algorithm. As the information showed in Fig. \ref{fig:week_mean_total_robustness} and Fig. \ref{fig:month_mean_total_robustness}, the mean value of the \textit{robustness} that got in a week or in a month is increased gradually when more polices involve into the patrolling. The figures of all the algorithms tend to converge at using 30 police officers. We also notice that G-LERK-CS, G-LERK-GA have better performance than their counterparts LERK-CS, LERK-GA  when the number of working officers is less than 15 people. However, the gap of the performance between algorithms is shrank when the number of polices who participated into the patrolling officers reaches 20 or 30. Moreover, from Fig. \ref{fig:week_mean_total_robustness} and Fig. \ref{fig:month_mean_total_robustness}, we can observe that the greedy algorithm performs better than the advanced GA and CS when the number of patrolling officers is less than 15 people. As greedy algorithm selects the next visiting node based on the highest patrol reward, it is more robust to the emergencies. However, when many patrolling officers are involved the system becomes more complex. In this scenario, GA and CS based algorithms outperform greedy based algorithms. Among two variants of greedy algorithm, Greedy-Imp outperforms Greedy-dist. It verifies that patrolling a node based on their possibility of crime event occurrence provide more robust solution than patrolling nearby nodes from current position.

\subsubsection{The Total Efficiency Obtained}
We also evaluate the weekly and monthly efficiency of the algorithms and illustrate them in Fig.~\ref{fig:week_mean_total_efficiency} and Fig.~\ref{fig:month_mean_total_efficiency} respectively. Similar to robustness calculation, 5-30 police officers are allocated in the missions and the results of total \textit{efficiency} each week are recorded and compared. The mean and standard deviation of the total \textit{efficiency} are calculated based on five tests for each algorithm. Both Fig.~\ref{fig:week_mean_total_efficiency} and Fig.~\ref{fig:month_mean_total_efficiency} indicate that the efficiency of the algorithms on solving the problem can be enhanced with more polices for the patrolling. In addition, the figures also demonstrate that G-LERK-CS and G-LERK-GA often perform better than their counterpart LERK-CS and LERK-GA respectively when the number of police officers is less than 20. But with the increasing number of officers, the former algorithms and the latter ones show a neck-and-neck result in the total efficiency. We also observe that when more police officers are involved, the greedy algorithm base solutions are more efficient in catching crime than the solutions obtained from GA and CS algorithms. The reason behind such result is when more police officers are involved they cover many observation areas. Therefore, visiting nearby locations only help them deterring crime event efficiently. The overall results provide us an insight that we could choose a suitable implementation to solve the proposed issue according to available human resources and the management effect worth more attention, such as robustness or efficiency.  
\section{Conclusion}\label{sec:conclusion}
This work focus on designing effective police surveillance strategy to prevent the future crime event occurrences and response to the emergency events. Compared with the existing studies on police patrol route designing where the potential nodes are selected based on spatial segmentation, this work include crime event prediction method where human mobility is considered as environment sensor to determine the next potential hotspot nodes. The provided model is also responsive to the uncertain arrival of emergency situations that is required to attend by a police officer. The findings in this research suggest that by fusing crime event prediction and real-time emergencies, the planning of patrol routes for multiple officers can be done more effectively.



\ifCLASSOPTIONcaptionsoff
  \newpage
\fi



%
\bibliographystyle{IEEEtran}
\bibliography{IEEEabrv,citation}
\end{document}